\documentstyle[12pt]{article}
\begin{document}
\title{\bf THE ELEMENTARY p(p,p$'\pi^{+}$)n REACTION\\}

\author{\bf Bijoy Kundu, B. K. Jain and A. B. Santra \\
Nuclear Physics Division \\
Bhabha Atomic Research Centre, Mumbai-400 085, India.\\}
% make date line blank
%\date{}

\maketitle

\begin{abstract}
A detailed study of the elementary p(p,p$'\pi^{+}$)n reaction is presented
using the delta isobar model. In this model, in the first step one of 
the two protons in the initial 
state gets excited to $\Delta $. This, in the second step, decays into a
nucleon and a pion. For the pp$\rightarrow$N$\Delta $ step the parametrized
form of the DWBA t-matrix of Jain and Santra, which reproduces most of 
the available data on pp$\rightarrow$n$\Delta ^{++}$, is used.  
The cross-sections studied include
the outgoing proton momentum spectra in coincidence with the pion, the
outgoing pion momentum spectra and the integrated total cross-section.
We find that all the calculated numbers are in good agreement with the 
corresponding measured cross sections.
\end{abstract}
PACS number(s): 25.40.Ve, 13.75.-n, 25.55.-e
\newpage

\section{\bf Introduction}
In past various authors \cite{Dmi-86,Jain-90}, 
including two of the present authors 
(Jain and Santra), have analysed theoretically the data on the 
 pp$\rightarrow $n$\Delta ^{++}$ reaction to extract the 
 potential for pp$\rightarrow $N$\Delta $ transition. In them, 
 the calculations 
of Jain et al. \cite{Jain-90} were done in the DWBA and those of 
Dmitriev \cite{Dmi-86} were done in the PWBA. 
 They concluded 
that the spin averaged data on the pp$\rightarrow $N$\Delta $ reaction 
can be reproduced very well by a one pion-exchange potential with the 
length parameter $\Lambda _\pi $ around 1-1.2 GeV/c in DWBA and around 
650 MeV/c in the PWBA. The difference in the two values of $\Lambda _\pi $  
is due to distortion effects. In fact, 
subsequently, when Jain et al.   
parametrized their DWBA t-matrix \cite{Jain-92}, 
they found that
the imaginary part of this t-matrix is very weak and the real 
part resembles to 
a great extent the one pion-exchange potential, with  
$\Lambda _\pi $ reduced to around 650 MeV/c.
 
The experimental data which above studies used were somewhat 
inclusive   
\cite{Shim,Bugg}. They were deduced from the 
pp$\rightarrow $np$\prime \pi^+$ reaction data which did not have  
the complete exclusive kinematics. The delta was 
identified in them by seeing a bump
in the missing mass spectrum.
A kinematically complete data set, however, exists on the 
pp$\rightarrow $p$\prime \pi ^+$n 
reaction at 800 MeV beam energy from LAMPF due to Hancock et al.\cite{Han}. 
They are a good coincidence data, and, thus, 
provide an excellent opportunity to test in detail 
the correctness of the  pp$\rightarrow $n$\Delta ^{++}$ DWBA t-matrix 
developed by two of us earlier \cite{Jain-92}. In the present paper we 
analyse the LAMPF data using this t-matrix. This
includes the analysis of the various  proton and pion energy
spectra measured in coincidence 
and the total integrated cross section for the 
pp$\rightarrow $p$\prime \pi ^+$n reaction.
 We assume that  
the pp$\rightarrow $p$\prime \pi ^+$n reaction proceeds 
in two steps. In the first step, one of the protons 
in the entrance channel gets converted to $\Delta $, and in the 
second step this delta 
decays into a pion and a nucleon. The transition matrix 
for the pp$\rightarrow \Delta $N step is taken to be the DWBA t-matrix 
mentioned above. The decay of the 
 delta is described by the pseudovector 
non-relativistic Lagrangian,
\begin {equation}
L_{\pi N\Delta}=i\frac {f^*_\pi }{m_\pi}({\bf S.\kappa _\pi})({\bf T.\phi }),
\end {equation}
where $f^*_\pi $ is the coupling constant at the $\pi $N$\Delta $ vertex. 
${\bf S}$ and ${\bf T }$ are the spin and isospin transition operators, 
respectively. This framework for the pp$\rightarrow $p$\pi ^+$n reaction  
includes in a certain way the final state 
interaction [FSI] amongst p$\pi ^+$n in the final state. The FSI 
consists of the interaction
between p and $\pi ^+$ and between the p$\pi ^+$ pair and the 
recoiling neutron. 
The dominant effect of the interaction between p and $\pi ^+$ is to
produce the $\Delta ^{++}$ resonance. This is explicitely included in our 
framework. The interaction between p$\pi ^+$ and the neutron in our
framework is approximated by that between the $\Delta ^{++}$ and the
neutron. A recent work by Jain and Kundu \cite{Jain-96} on the delta decay 
in nuclear 
medium suggests that this approximation is reasonably good.  

The pp$\rightarrow$np$'\pi^{+}$
process has also been worked out in the literature by
Engel et. al \cite{Shyam}.
However, these calculations use plane waves for the continuum particles.
Thus, unlike our work, this work
does not include the effect of distortions in the
entrance and the exit channels.

Inclusion/omission of rho-exchange in 
the description of the  pp$\rightarrow $n$\Delta ^{++}$ reaction has been 
the topic of much debate in the literature. 
The general conclusion is that the spin averaged data on the 
pp$\rightarrow  \Delta ^{++}$n reaction are well reproduced by 
one pion-exchange potential only 
\cite{Wick,Dmi-86,Jain-90,Jain1}.  
Any attempt 
 to include the rho-exchange worsens the agreement with the experiments, 
 and yield unsatisfactory results. 
 In this context it is also interesting
 to see the work of Jain et al. \cite{Jain2} 
 which discusses the relative importance of rho-exchange in  
 p(n,p)n and p(p,n)$\Delta ^{++}$ reactions. They conclude that, while it 
 is absolutely essential to include the rho-exchange in the 
 description of the p(n,p)n reaction, the rho-exchange is not required
 for accounting the p(p,n)$\Delta ^{++}$ data. This study deals with the 
 spin averaged cross sections. A recent theoretical study on the
 microscopic structure of the $\rho N\Delta $ vertex by Haider et al.
 \cite{Haider} supports this conclusion. They find 
 that the microscopically calculated value of the f$_{\rho N \Delta }$ 
 coupling constant is much 
 smaller than what is normally assumed.
 The measured spin averaged cross sections on nuclei in charge
 exchange reactions are also reproduced with only a pion exchange \cite{Dmi1}.
 It is, however, true that the measurements of Prout et al.
\cite{Prout} with a polarized proton beam on 
 nuclei, and earlier by Ellegaard et al. \cite{Elle}
 do show a large transverse part. But, as shown by 
 V.F.Dmitriev \cite{Dmi1} and
 Sams et al.
 \cite{Sam}, large transverse contribution can 
 also arise from the distortion of the continuum particles.
All these discussions thus suggests that, at best, the role of 
 rho-exchange in the charge-exchange reaction in the delta region 
 is controversial. The spin averaged cross sections do not need it, 
 the spin transfer measurements show some indications for it. Since 
 the present work deals with the spin averaged cross sections, our use of one 
 pion-exchange is consistant with other work in this field. 

In section 2 we write the formalism for the pp$\rightarrow$np$'\pi^{+}$
process. Section 3 gives calculated cross sections for the proton and pion
energy spectra at 800 MeV beam energy 
and the total cross section from 500 MeV to 2 GeV. 
These results are compared with the available experimental cross sections.
A good agreement is obtained.

\section {\bf Formalism}
The cross-section for the
pp$\rightarrow$np$'\pi^{+}$ process is given by
\begin{equation}
d\sigma= <|(t_{pp\rightarrow p'\pi^+ n}|^2>
[PS],
\end{equation}
where, the angular brackets denote the sum and average over the 
spins in the initial and final
states, respectively.
[PS] is the factor associated with the phase-space and the beam
current. For the proton and pion detected in coincidence in the final 
state, in the lab. frame it is given by,
\begin{equation}
[PS]= \frac{m_p^2 m_n k_p'^2 k_\pi^3}
{2 (2\pi)^5
k_p E_p'} \frac{1}{k_\pi^2 (E_i-E_p')-E_\pi |{\bf(k_p-k_p').k_\pi |}}
d\Omega_p' d\Omega_\pi dk_p'.
\end{equation}

$t_{pp\rightarrow p'\pi^+ n}$ is the t-matrix for the 
$pp\rightarrow p'\pi^+ n$ process. It consists of two parts: 
one corresponding to the excitation of the proton in the initial 
state to  $\Delta^{++}$ and another corresponding to its 
excitation to $\Delta^{+}$ [Figure 1]. That is 
\begin{equation}
t_{pp\rightarrow p'\pi^+ n}=t^{\Delta^{++}}+t^{\Delta^{+}}.
\end{equation}
Furthermore, because of the antisymmetrization of the protons, each
t-matrix in turn consists of two terms, one corresponding to the 
excitation of the beam proton and another corresponding to the 
excitation of the target proton. We call them ``direct'' and ``exchange"
terms, respectively.

Putting every thing together, we get
\begin{eqnarray}
t_{pp \rightarrow NN \pi} & = & \sum _{\Delta}
<N \pi| {\bf S.\kappa_\pi}
{\bf T.\phi_\pi}|\Delta> \nonumber\\
&\times& G_\Delta <t_{pp\rightarrow N\Delta}>,
\label{ttmat}
\end{eqnarray}
where  N represents
a proton or a neutron in the final state
corresponding to the decay of 
$\Delta^{++} \rightarrow \pi^{+}$p and $\Delta^{+}
\rightarrow \pi^{+}$n, respectively.
$\Delta$ stands for a $\Delta^{++}$ or $\Delta^{+}$
excitation in the intermediate state.
$\kappa_\pi$ at the $\Delta$-decay vertex is the
outgoing pion momentum in the $\pi$N
centre-of-mass.
It is given by,
\begin{eqnarray}
\kappa_\pi (\mu^2,m_\pi^2)=[(\mu^2+m^2-m_\pi^2)^2/4\mu^2-m^2]^{1/2}.
\end{eqnarray}
This relation reflects the restrictions on the available phase space for the
decay of a delta of mass $\mu$ into an on-shell pion of mass m$_\pi$ 
(=140 MeV) and a nucleon.
Since the final outgoing pion is on-shell, the $\Delta$N$\pi$ vertex
does not contain the usual
form factor F$^{*}$. $G_\Delta$ in equation \ref{ttmat}
is the delta propagator. Its form is
taken as,
\begin{eqnarray}
G_\Delta= \frac{2 m_\Delta}{\mu^{2}-m_\Delta^{2}+i\Gamma_\Delta m_\Delta},
\end{eqnarray}
where, m$_\Delta$(=1232 MeV) and $\Gamma_\Delta$ are the resonance parameters
associated with a free $\Delta$. The free width, $\Gamma_{\Delta}$,
depends upon the
invariant mass and is written as,
\begin{eqnarray}
\Gamma _\Delta=\Gamma_0 \Bigl[\frac{ k(\mu^2,m_\pi^2)}
{k(m_\Delta^2,m_\pi ^2)}\Bigr]^3
\frac{k^2(m_\Delta ^2,m_\pi ^2)+\gamma ^2}{k^2(\mu ^2,m_\pi ^2)+\gamma ^2},
\label{freewidth1}
\end{eqnarray}
with $\Gamma _0$=120 MeV and $\gamma$=200 MeV. $\mu$ is the invariant mass
of the N$\pi^{+}$ system and is given by,
\begin{eqnarray}
\mu ^2=(E_{N}+E_\pi)^2-({\bf k}_{N}+{\bf k}_\pi)^2 .
\label{inv}
\end{eqnarray}

$t_{pp\rightarrow N\Delta}$ is the DWBA t-matrix for the 
$pp\rightarrow N\Delta$ transition. Following Jain and Santra \cite{Jain-90}, 
it is given by
\begin{eqnarray}
t_{pp\rightarrow N\Delta}= 
(\chi^{-}_{\bf k_f}, <n \Delta^{++}|v_\pi|
\{pp\}>,\chi_{\bf k_i}^{+}),
\label{disttmat}
\end{eqnarray}
where curly brackets around pp represent the antisymmetrization
of the pp wave function. $v_\pi $ is the one pion-exchange potential for
$pp\rightarrow N\Delta$ transition.
 $\chi$s are the distorted waves. They describe the elastic
scattering of the pp and the n$\Delta$ systems.
Jain and Santra [4] have evaluated equation \ref{disttmat} using
eikonal approximation for $\chi$s.
With $\Lambda
_{\pi}$=1 GeV/c at both the $\pi$NN and $\pi$N$\Delta$ vertices,
they found that this t-matrix reproduces the available experimental data
on this reaction over a large energy range very well.

Jain and Santra also found that their DWBA 
t-matrix can be easily parametrized
\cite{Jain-92}. The parametrized t-matrix is complex, but 
its imaginary part is very weak. The real part resembles very much 
with the one pion-exchange potential with its  length parameter, 
$\Lambda _\pi $, reduced to around 600-700 MeV/c. For the present
calculations, instead of repeating the full calculation of the t-matrix,
 we have used the parametrized form, i.e.  
\begin{eqnarray}
t_{pp\rightarrow N\Delta } \approx
\it v_\pi ^{pp\rightarrow N\Delta } (\Lambda_\pi=650 MeV/c)= -\frac {ff^*}
{m_\pi^2}
FF^*  \frac {\bf S^+ . q \sigma . q}{m_\pi^2 +q^2-\omega^2} {\bf T^+.\tau},
\label{vpot}
\end{eqnarray}
where f and f$^*$ at the $\pi$NN and $\pi$N$\Delta$ vertices are 1.008
and 2.156 respectively \cite{Bug1}.
${\bf q}$ is the momentum transfer in the pion-nucleon rest frame. 
Since the exchanged pion is virtual, it is not straight forward 
to define this 
momentum quite unambiguously.   
For the $\pi$N$\Delta$ vertex we use the following
Galilian invariant form,
\begin{eqnarray}
{\bf q}= {\bf k_p -k_\Delta [=(k_N+k_\pi)]}-
\frac{\omega {\bf k_\Delta}}{E_\Delta},
\label{pinn}
\end{eqnarray}
where, $\omega$ is the energy transfer in exciting the $\Delta$.
At the $\pi$NN vertex we replace
\begin{eqnarray}
{\bf q}^2 \rightarrow {-t},
\label{pindel}
\end{eqnarray}
where t is the four momentum squared.
\section{Results and Discussion}
Using the above formalism we calculate
the exclusive proton momentum spectra,
the outgoing pion momentum spectra
and the integrated total p(p,p$'\pi^{+}$)n cross-section.

As the detailed measurements for the p(p,p$'\pi^{+}$)n process exist at
800 MeV beam energy, we first calculate the differential cross-sections
at this energy.
In figure 2, we plot the calculated as well as the measured \cite{Han} 
exclusive proton momentum spectra for
the proton and the pion angles of 14.5$^0$ and
-21$^0$ degrees , respectively. These angles correspond to the 
delta going at
0$^0$.
The figure has got four calculated curves.
The short-dashed and dot-dashed curves correspond to $\Delta ^{++}$ and 
and $\Delta ^+$ contributions (including both, the 
``direct" as well as ``exchange"  diagrams), respectively. 
The solid curve is the 
coherent sum of these two contributions. We find that this curve agrees 
well with 
the measured cross sections. We also note that the 
main contribution to the solid curve comes from the $\Delta ^{++}$ 
diagram. The   
$\Delta ^+$ contributes  only to the extent of 
5-10 $\%$.  

To show the contribution of the ``exchange" diagram, in fig. 2 we also
show ( by
long-dashed curve) the cross section 
 for the $\Delta ^{++}$ diagram using only 
the ``direct" term. Comparing this with the
short-dashed curve, which includes both the direct and exchange
diagrams, we find that the contribution of the exchange term is 
around 15-20$\%$.

In figure 3, we show the proton spectrum
for another set of proton and pion angles. This pair of angles also correspond
to the delta going at 0$^0$.
The
outgoing proton and pion angles are 
 14.5$^0$ and -42$^0$, respectively.
All the curves have the same meaning as those in figure 2.
Here too the calculated proton spectrum is in good accord with the
measured spectrum. Other observations also remain same as in fig. 2.

In figure 4
we show the double differential cross-section as a function
of the outgoing pion momentum. The proton angles are integrated.
Experimentally such measurements exist for 800 MeV beam energy and 
the pion detected at 20$^0$
\cite{Bev}.
In this figure we have 3 curves alongwith the experimental data. 
The dash and dash-dot curves  correspond 
separately to the $\Delta^{++}$ and $\Delta^{+}$
diagrams, respectively. The solid curve is calculated 
including both the diagrams. All the curves include the direct
as well as exchange diagrams. Excluding the peak in the measured 
cross sections around 550 MeV, the solid curve is in overall accord with 
the measured cross sections.  Relative contributions of the
 $\Delta ^+$ and $\Delta ^{++}$ to the cross sections are at the same 
 level as in the earlier curves. 
 The peak around 550 MeV, as kinematic 
considerations suggest, may arise from the resonance structure  between
 neutron and proton in the final state.

Finally in figure 5 we present the calculated
total integrated cross section as a function
of the beam energy from threshold to 2 GeV. Since, as seen from the results
in figures 2 - 4, the contribution of the
$\Delta^{+}$ is only at the level of 10$\%$, we give the calculated results
for the $\Delta^{++}$ only. The calculated results include both the
direct and the exchange contributions.
We find an excellent agreement between the calculated and measured
cross-sections \cite{Lock}.
\newpage
\section {Conclusions}
In conclusion, the findings of this paper can be summarized as :
\begin{enumerate}
\item Experimentally measured exclusive proton momentum spectra, 
the pion momentum
spectrum and the total integrated cross sections over a large energy 
range can be reproduced well with
one-pion exchange potential for the delta excitation in the 
intermediate state;
\item the contribution of the $\Delta^{++}$ dominates.
$\Delta^{+}$ contributes only to the extent of
5-10$\%$, and
\item the effect of the exchange
process is to bring down the cross-section.
Its contribution, however, is
only at the level
10-20$\%$.
\end{enumerate}

\newpage
%\begin{references}
\section *{References}

%\end{references}

\newpage
{\bf Figure Captions}
\begin{enumerate}
\item The direct and exchange diagrams for the $\Delta $ excitation.
\item The outgoing proton momentum spectrum in coincidence with the pion.
T$_p$=800 MeV.
$\theta_p'$=14.5$^0$ and
$\theta_\pi$=-21$^0$.
The experimental points are from \cite{Han}.
The long-dashed curve is calculated using the direct $\Delta^{++}$
diagram and the short-dashed curve includes both the direct and the
exchange $\Delta^{++}$ diagrams. The solid curve is calculated
using both the $\Delta^{++}$ and $\Delta^{+}$ diagrams added coherently.
The dash-dot curve is the $\Delta^{+}$ contribution
multiplyied by 5.
$\Lambda_\pi$=650 MeV/c.

\item Same as figure 2 with
$\theta_p'$=14.5$^0$ and
$\theta_\pi$=-42$^0$. Experimental points are from \cite{Han}.
All the curves have the same meaning as in figure 2.
$\Lambda_\pi$=650 MeV/c.
\item The outgoing pion momentum spectra for the p(p,p$'\pi^{+}$)n reaction
at T$_p$=800 MeV.
$\theta_\pi$=20$^0$.
The experimental points are from \cite{Bev}.
The solid curve is calculated using both the $\Delta^{++}$ and $\Delta^{+}$
diagrams added coherently. The short-dashed and dot-dashed curves 
show separately the contribution due to 
$\Delta^{++}$ and $\Delta^{+}$, respectively.
$\Lambda_\pi$=650 MeV/c.
\item Total cross-section for the p(p,p$'\pi^{+}$)n reaction. The
calculated curve includes both the direct and exchange $\Delta^{++}$
excitation diagrams.
$\Lambda_\pi$=650 MeV/c.
The experimental points are from \cite{Lock}.
\end{enumerate}
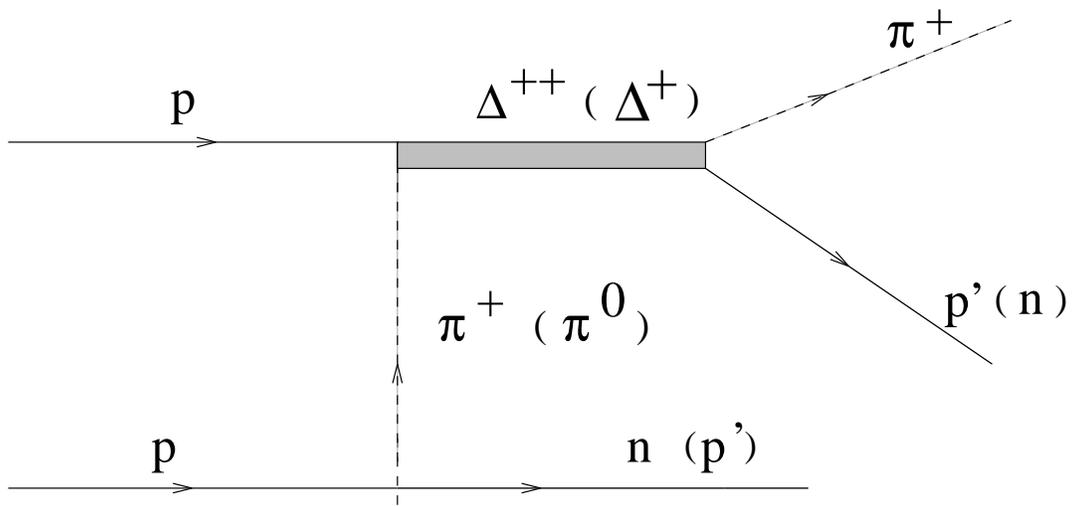
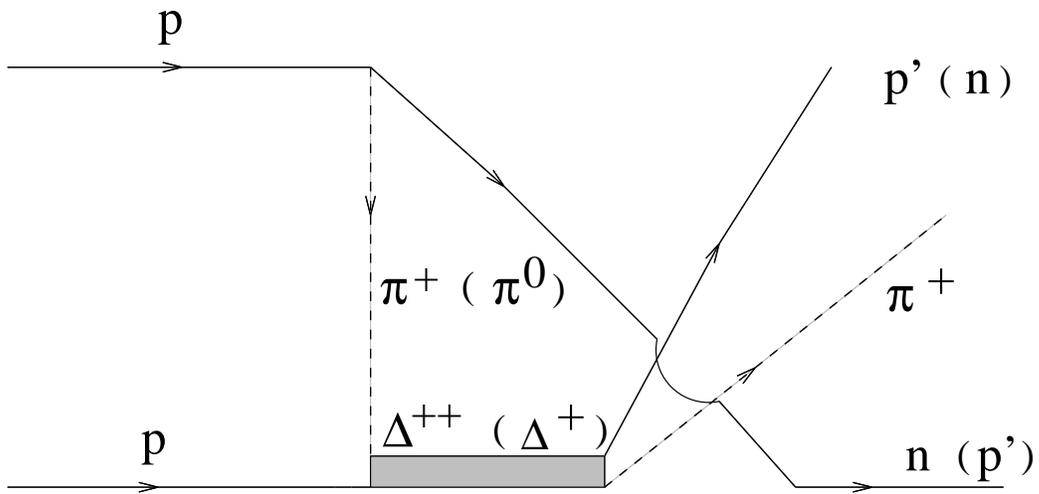
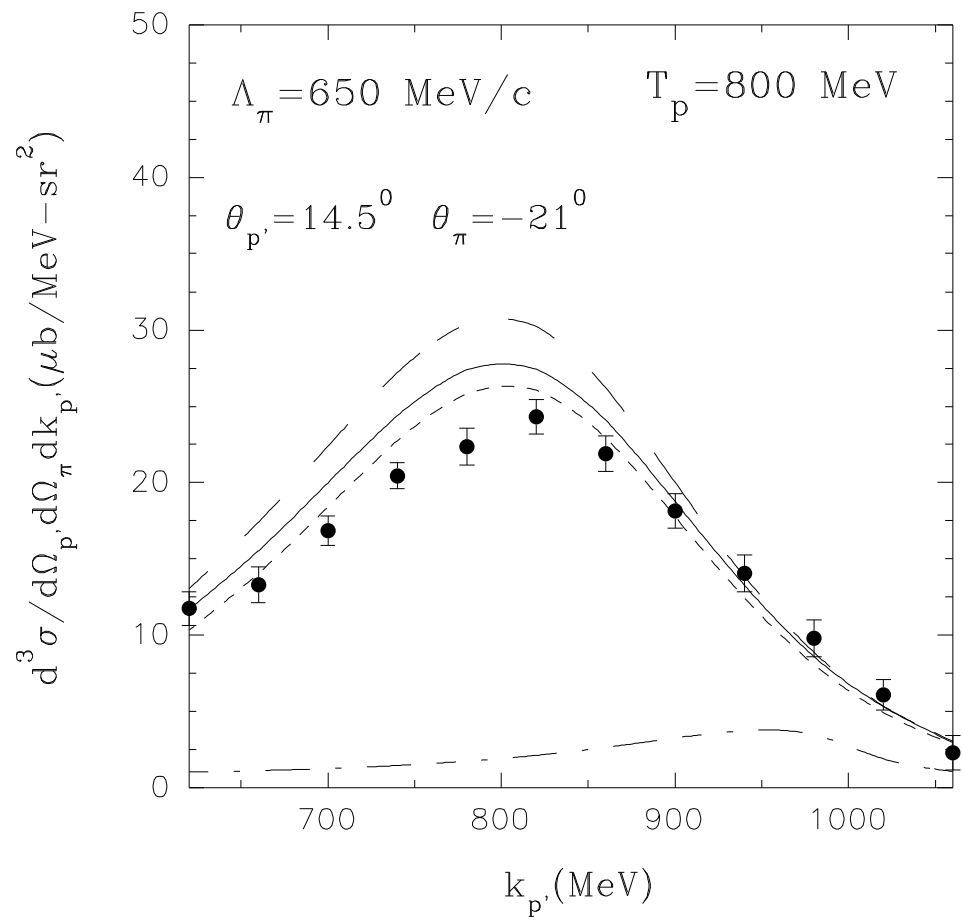
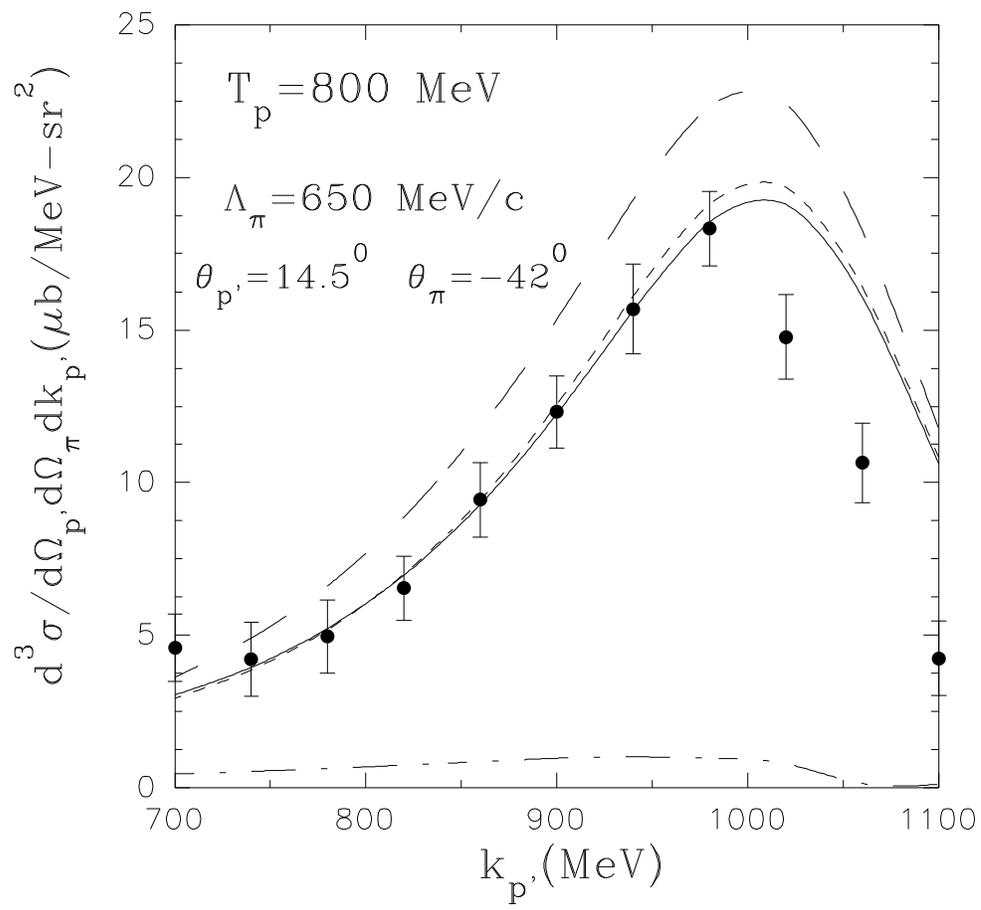
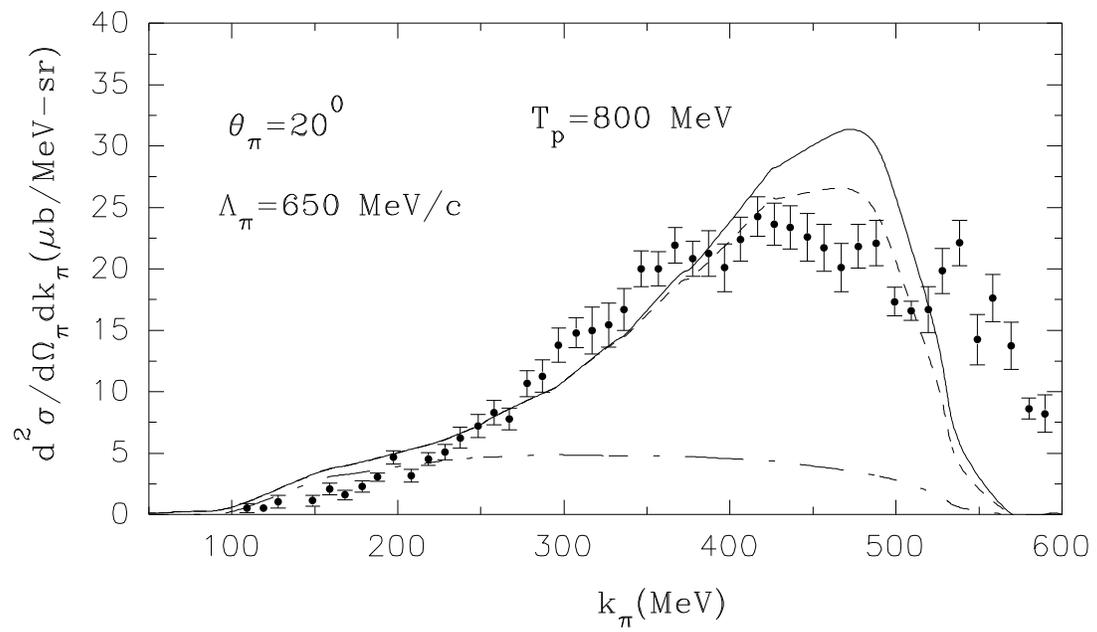
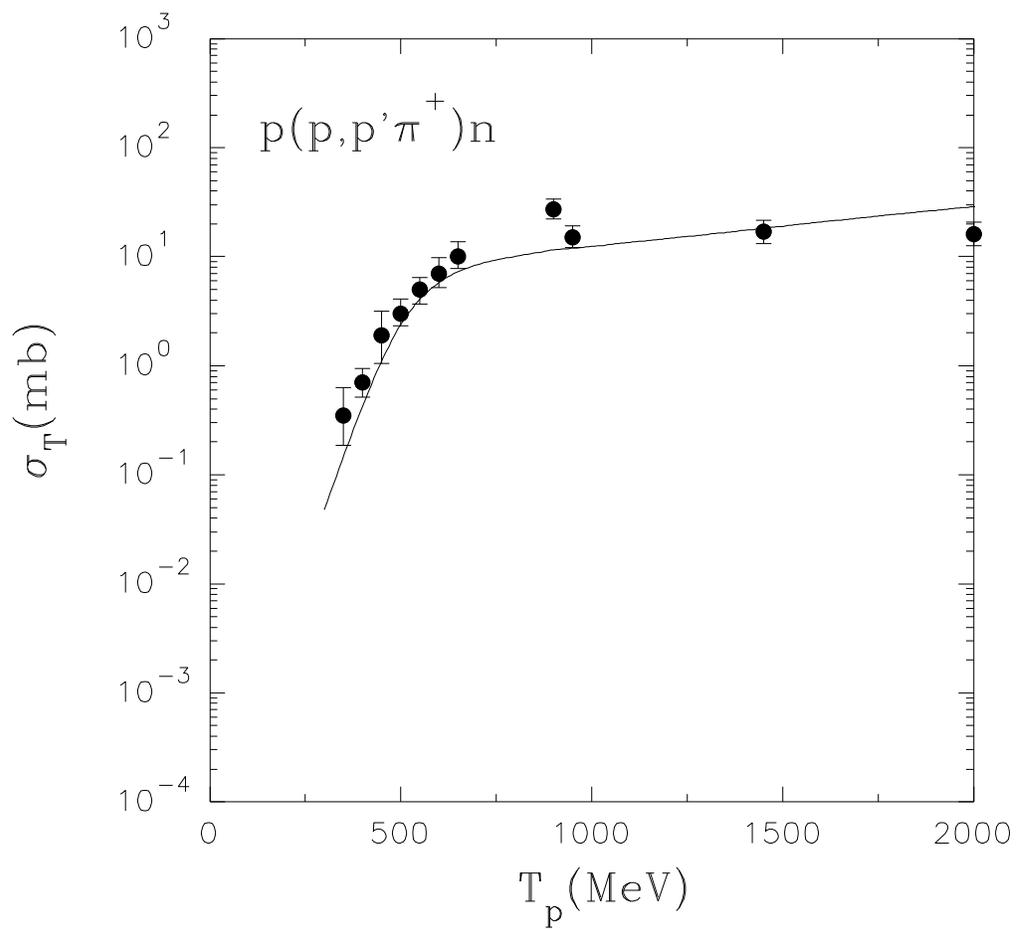
\end{document}